\documentclass{article}      

\usepackage{epsfig}

\title{q-Shock Soliton Evolution}

\author{Oktay K. Pashaev and  Sengul Nalci \\Department of Mathematics, Izmir Institute of Technology \\ Urla-Izmir, 35430, Turkey}

\begin{document}

\maketitle


\begin{abstract}
  By generating function based on the Jackson's q-exponential function and standard exponential function, we introduce a new q-analogue of Hermite and Kampe-de Feriet polynomials.
In contrast to standard Hermite polynomials, with triple recurrence relation, our polynomials satisfy multiple term recurrence relation, derived by the q-logarithmic function. It allow us to introduce the q-Heat equation with standard time evolution and the q-deformed space derivative. We found solution of this equation in terms of q-Kampe-de Feriet polynomials with arbitrary number of moving zeros, and solved the initial value problem in operator form. By q-analog of the Cole-Hopf transformation we find  a new q-deformed Burgers type nonlinear equation with cubic nonlinearity. Regular everywhere single and multiple q-Shock soliton solutions and their time evolution are studied. A novel, self-similarity property of these q-shock solitons is found. The results are extended to the time dependent q-Schr\"{o}dinger equation and the q-Madelung  fluid type representation is derived.

\end{abstract}

\section{Introduction}
It is well known that the Burgers' equation in one dimension can be reduced via the Cole-Hopf transformation to the linear heat equation.
It allows one to solve the initial value problem for the Burgers equation and get exact solutions in
the form of shock solitons and their scattering. In the present paper we introduce the differential-q-difference Burgers type  equation with cubic nonlinearity which includes the standard time derivative and the q-deformed space derivative.By using the q-Cole-Hopf transformation, this nonlinear equation can be linearized in terms of the q-Heat equation with standard time evolution and q-different space derivative. Based on the Jackson's q-exponential function and the standard exponential function we introduce a new q-analog of Hermite and Kampe-de Feriet polynomials, representing moving poles solution for the q-Burgers equation. Then we derive the operator solution of the initial value problem (IVP) for the q-Burgers
equation in terms of the IVP for the q-heat equation. We construct several particular solutions of our q-Burgers type equation in the form of regular everywhere q-shock solitons with regular time evolution free of singularities. It turns out
that the static q-shock soliton solution of our equation shows remarkable self-similarity property in space coordinate $x$. By extending our results to the complex domain we introduce the time dependent Schr\"{o}dinger equation with q-deformed dispersion and the complex wave function. As a solution of this equation we get the set of complex q-Kampe-de Feriet polynomials. By the complex q-Cole-Hopf transformation we obtain the complex q-Burgers-Madelung equation as a coupled two fluid system with complex velocity function.

\section{q-Hermite Polynomials}

 We define a q-analog of Hermite polynomials by the generating function with the Jackson's q-exponential function \cite{Jackson} and standard exponential function as \begin{equation}e^{-t^2} e_q([2]_q t x) = \sum^{\infty}_{N=0} H_N(x;q) \frac{t^N}{[N]_q!},\label{genfunc}\end{equation}
where the Jackson's q-exponential function is defined by $$e_q(x)= \sum^{\infty}_{n=0} \frac{x^n}{[n]_q!},$$  $[n]_q!=[1]_q[2]_q...[n]_q$ and q-number \cite{Kac et al.},
$$[n]_q= \frac{q^n-1}{q-1}.$$

From the defining identity (\ref{genfunc}) it is not difficult to derive for the q-Hermite polynomials an explicit sum formula

\begin{equation}H_N(x; q)=  \sum^{[N/2]}_{k=0} \frac{(-1)^k ([2]_q x)^{N-2 k} [N]_q!}{k! [N-2 k]_q! }.
 \end{equation}
This explicit sum makes it transparent in which way our polynomials $H_N(x;q)$ q-extended the $H_N(x)$ and how they are different from the known ones in literature.
By $q$-differentiating the generating function (\ref{genfunc}) with respect to x we derive the two-terms recurrence relation
\begin{equation} D_x H_N(x;q)= [2]_q [N]_q H_{N-1}(x;q), \label{rec2} \end{equation} where the q-derivative is defined as \cite{Kac et al.}
\begin{equation}D_x f(x)=\frac{f(qx)-f(x)}{(q-1)x}.\end{equation}

And by standard differentiating the generating function (\ref{genfunc}) with respect to t and using the next evident equality
$$ t \frac{d}{dt} e_q([2]_q x t)= x \frac{d}{dx} e_q([2]_q x t)= \sum^{\infty}_{n=0} n \frac {([2]_q x t)^n}{[n]_q!}$$ we obtain another two-term recurrence relation
\begin{equation} (x \frac{d}{dx} - N) H_N(x;q)= 2 [N]_q [N-1]_q H_{N-2}(x;q). \label{rec3}
\end{equation}
 By standard differentiating the generating function (\ref{genfunc}) with respect to t and using  definition of the q-logarithmic function \cite{Pashaev et al.}
 $$ Ln_q(1+z)= \sum^{\infty}_{N=1} \frac{(-1)^{N-1} z^N}{[N]},$$ where $q>1, 0<|z|<q$ and the property $$\frac{d}{dz} Ln \,\,e_q \left(\frac{\alpha z}{1-q}\right)=\frac{ Ln_q(1-\alpha z)}{(q-1)z} $$
we derive the N-term recurrence relation formula
\begin{eqnarray}&&H_{N+1}(x;q)= \frac{[N+1]_q}{N+1}\{ [2]_q x H_N(x;q)-2[N]_q H_{N-1}(x;q)-  \\\nonumber  \\&&(q-1)[2]_q [N]_q x^2 H_{N-1}(x;q)+ [2]_q [N]_q! \sum^{N-2}_{k=0}
\frac{(1-q^2)^{N-k} x^{N-k+1} H_k(x;q)}{[k]_q! [N-k+1]_q}\}\nonumber \end{eqnarray}
When $q\rightarrow 1$ this multiple term recurrence relation for q-Hermite polynomials reduces to the three-term recurrence relation for the standard Hermite polynomials
$$H_{N+1}(x)=2 x H_N(x)-2 N H_{N-1}(x).$$

Substituting (\ref{rec2}) into N-term recurrence relation formula we get the operator representation
\begin{eqnarray}&&H_{N+1}(x;q)=\\&&\frac{[N+1]_q}{N+1}\left( [2]_q x-(\frac{2}{[2]_q}+(q-1)x^2)D_x+\sum^{N}_{l=2}\frac{(1-q^2)^l x^{l+1}}{[2]_q^{l-1}[l+1]_q}D_x^l\right)H_N(x;q) \nonumber \end{eqnarray}
By the recursion, starting from $n=0$ and $H_0(x;q)=1$ we have next representation for the q-Hermite polynomials
\begin{eqnarray}&&H_{N+1}(x;q)=\\&& \prod^{N}_{k=0}\frac{[k+1]_q}{k+1}\left( [2]_q x-(\frac{2}{[2]_q}+(q-1)x^2)D_x+\sum^{N}_{k=2}\frac{(1-q^2)^k x^{k+1}}{[2]_q^{k-1}[k+1]_q}D_x^k\right)\cdot 1 \nonumber
\end{eqnarray}
In the limit  $q\rightarrow 1$ case this product formula is reduced to the known one
$$H_N(x)=(2 x- \frac{d}{d x})^N \cdot 1 $$

 We note that the generating function and   the form of our q-Hermite polynomials are different from the known ones in the literature, \cite{Exton}, \cite{Cigler et
al.},
\cite{Rajkovic et al.}, \cite{Negro}, \cite{Nalci et al.}. Moreover, instead of three-term recurrence relation we have multiple term recurrence relation, which shows that our q-Hermite polynomials are different from the known ones for orthogonal polynomial sets \cite{Ismail}.

 The first few q-Hermite polynomials are
\begin{eqnarray} H_0 (x;q) = 1\nonumber ,\end{eqnarray}\begin{eqnarray}H_1 (x;q) = [2]_q \,x \nonumber,\end{eqnarray}
\begin{eqnarray} H_2(x;q) = [2]^2_q \, x^2 - [2]_q\nonumber,\end{eqnarray}
\begin{eqnarray} H_3(x;q) = [2]^3_q \,x^3 - [2]^2_q [3]_q \,x\nonumber, \end{eqnarray}
\begin{eqnarray} H_4(x;q) = [2]^4_q \,x^4 - [2]^2_q [3]_q [4]_q\,x^2 + \frac{1}{2} [4]_q!\nonumber .\end{eqnarray}
When $q\rightarrow 1$ these polynomials reduce to the standard Hermite polynomials.

\subsection{q-Difference Equation}

Applying $D_x$ to both sides of  (\ref{rec2}) and using recurrence formula (\ref{rec3}) we
get the q-difference-differential equation for the q-Hermite polynomials
\begin{equation} 2 D^2_q H_N(x;q)-[2]^2_q x \frac{d}{dx} H_N(x;q)+[2]^2_q N H_N(x;q)=0 .\end{equation}

In $q \rightarrow 1$ limit it reduces to the second order linear differential equation for the standard Hermite polynomials $$\frac{d^2}{dx^2}H_N(x)-2 x \frac{d}{dx}H_N(x)+N H_N(x)=0.$$
\section{q-Kampe-de Feriet Polynomials}

We define the q-Kampe-de Feriet polynomials as
\begin{equation}
H_N(x, \nu t;q)= (-\nu t)^{\frac{N}{2}} H_N\left( \frac{x}{[2]_q \sqrt{-\nu t}}; q\right)
,\end{equation}
so that from N-term recurrence relation for q-Hermite polynomials we obtain N-term recurrence relation formula for q-Kampe-de Feriet polynomials
$$\begin{array}{l}H_{N+1}(x,\nu t;q)= \frac{[N+1]_q}{N+1}[ x H_N(x,\nu t;q)+2\nu t[N]_q H_{N-1}(x,\nu t;q)\nonumber \\  \\ -\frac{1}{[2]_q }(q-1)[N]_q x^2 H_{N-1}(x,\nu t;q)+  [N]_q! \sum^{N-2}_{k=0}
\frac{(1-q^2)^{N-k} x^{N-k+1} H_k(x,\nu t;q)}{[k]_q! [N-k+1]_q [2]_q^{N-k}}] \end{array}$$
  This can also be written in the operator form as
\begin{eqnarray}&&H_{N+1}(x,\nu t;q)=\\ && \frac{[N+1]_q}{N+1}\left[x+(2 \nu t+\frac{1-q}{[2]_q}x^2)D_x+\sum^{N}_{l=2}\frac{(1-q^2)^l x^{l+1}}{[2]_q^l [l]_q}D_x^l\right]H_N(x,\nu t;q)\nonumber \end{eqnarray}
By the recursion, starting from $n=0$ and $H_0(x,\nu t;q)=1$ we have next operator representation for the q-Kampe-de Feriet  polynomials
 \begin{eqnarray} &&H_{N+1}(x,\nu t;q)=\\ &&\prod^{N}_{k=0}\frac{[k+1]_q}{k+1}\left[x+(2 \nu t+\frac{1-q}{[2]_q}x^2)D_x+\sum^{N}_{k=2}\frac{(1-q^2)^k x^{k+1}}{[2]_q^k [k]_q}D_x^k\right]\cdot 1 \nonumber \end{eqnarray}
In $q \rightarrow 1$ case we have $$H_N(x)=(x+2 \nu t \frac{d}{d x})^N \cdot 1.$$

Then the first few q-Kampe-de Feriet polynomials are
\begin{eqnarray} H_0 (x, \nu t;q) = 1\nonumber,\end{eqnarray} \begin{eqnarray}H_1 (x, \nu t ;q) = x\nonumber,\end{eqnarray}
\begin{eqnarray} H_2(x, \nu t ;q) =  x^2 + [2]_q \, \nu t\nonumber,\end{eqnarray}
\begin{eqnarray} H_3(x, \nu t ;q) = x^3 + [2]_q [3]_q \,\nu t \,x\nonumber ,\end{eqnarray}
\begin{eqnarray} H_4(x, \nu t ;q) = x^4 + [3]_q [4]_q\,\nu t \,x^2 + \frac{[4]_q!}{2} \nu^2 t^2\nonumber.\end{eqnarray}
When $q\rightarrow 1$ these polynomials reduce to the standard Kampe-de Feriet polynomials.

\section{q-Heat Equation}
We introduce the q-Heat equation  \begin{equation}(\partial_t - \nu D^2_x) \phi(x,t) = 0
 \label{qheatstandardt}\end{equation} with partial q-derivative with respect to $x$  and with partial standard derivative in time $t$.
 One can easily see that $$\phi_k (x,t)=e^{\nu k^2 t} e_q (k x)$$ is a plane wave solution of (\ref{qheatstandardt}).
By expanding this in terms of parameter $k$
\begin{equation}\phi_k (x,t)=e^{\nu k^2 t} e_q (k x)= \sum^{\infty}_{N=0} H_N(x,\nu t;q) \frac{k^N}{[N]_q!} \label{genfuncforkampe}\end{equation}
we get the set of q-Kampe-de Feriet polynomial solutions for the q-Heat equation (\ref{qheatstandardt}).
From the defining identity (\ref{genfuncforkampe}) is not difficult to  derive an explicit sum formula for the q-Kampe de Feriet polynomials

\begin{equation}H_N(x,\nu t; q)=  \sum^{[N/2]}_{k=0} \frac{(\nu t)^k x^{N-2 k} [N]_q!}{k! [N-2 k]_q!}.
 \end{equation}

\subsection{Operator Representation}

\newtheorem{prop}{Proposition}
\newtheorem{pf}{Proof}
\newtheorem{cor}{Corrollary}

\begin{prop}

\begin{equation} e^{-\frac{1}{[2]^2_q} D^2_x } e_q ([2]_q x t) = e^{-t^2} e_q ([2]_q x t)\label{prop1}.\end{equation}
\end{prop}

\begin{pf}
By q- differentiating the q-exponential function with respect to $x$
\begin{equation} D^n_x e_q ([2]_q x t) = ([2]t)^n e_q ([2]_q x t),\end{equation}
and combining then to the sum
\begin{equation} \sum_{n=0}^{\infty} \frac{a^n}{n!} D^{2n}_x\,\, e_q ([2]_q x t) =  \sum_{n=0}^{\infty} \frac{([2]_q t)^{2n} a^n }{n!}\,\, e_q ([2]_q x
t),\end{equation}
we have relation
\begin{equation} e^{a D^2_x}\,\, e_q ([2]_q x t) = e^{a ([2]_q t)^2}\,\, e_q ([2]_q x t).\end{equation}
By choosing $a = - 1/[2]_q^2$ we get the result (\ref{prop1}).
\end{pf}

\begin{prop}

\begin{equation}H_N (x; q) =  [2]^N_q e^{-\frac{1}{[2]^2_q} D^2_x } x^N \label{prop2}.\end{equation}
\end{prop}
\begin{pf} The right hand side of (\ref{prop1}) is the generating function for the q-Hermite polynomials (\ref{genfunc}). Hence, equating the coefficients of $t^n$ on both sides gives the result.
\end{pf}

\begin{cor} If function $f(x)$ is expandable to the formal power series $f(x) = \sum^\infty_{N=0} a_N x^N$ then
we have next q-Hermite series
\begin{equation} e^{-\frac{1}{[2]^2_q} D^2_x } f(x) = \sum^\infty_{N=0} a_N\frac{H_N(x;q)}{[2]_q^N}.\end{equation}
\end{cor}
\section{Evolution Operator}

Following similar calculations as in Proposition I we have the next relation
\begin{equation} e^{\nu t D^2_x} e_q (k x) = e^{ \nu t k^2} e_q ( k x).\end{equation}
The right hand side of this expression is the plane wave type solution of the q-heat equation (\ref{qheatstandardt}).
Expanding both sides in power series in $k$ and equating the coefficients of $k^N$ on both sides we get
 q-Kampe de Feriet polynomial solutions of this equation
\begin{equation} H_N(x,\nu t;q) = e^{ \nu t D^2_x } x^N\end{equation}
as solution of the I.V.P. with $\phi(x,0)=x ^N.$

Consider an arbitrary, expandable to the power series function $f(x) = \sum^\infty_{n=0} a_n x^n$, then the formal series
\begin{eqnarray} f(x,t) = e^{\nu t D^2_x } f(x) &=& \sum^\infty_{n=0} a_n  e^{\nu t D^2_x } x^n \\ &=& \sum^\infty_{n=0} a_n
H_N(x, \nu t;q),\end{eqnarray}
represents a time dependent solution of the q-heat equation (\ref{qheatstandardt})
The domain of convergency for this series is determined by asymptotic properties of our q-Kampe-de Feriet polynomials for $n\rightarrow \infty$
and requires additional study.

According to this we have the evolution operator for the q-heat equation as
\begin{equation} U(t) = e^{\nu t D^2_x }.\end{equation}
It allows us to solve the initial value problem
\begin{eqnarray}(\frac{\partial}{\partial t} - \nu D^2_x) \phi(x,t) &=& 0 ,\\
 \phi(x,0) &=& f(x),\end{eqnarray}
in the form
\begin{equation} \phi(x,t) =  e^{\nu t D^2_x } \phi(x,0) = e^{\nu t D^2_x } f(x),\label{IVP}\end{equation}
where we imply the base $q>1$ so that $e_q(x)$ is an entire function.

\section{q-Burgers' Type Equation}

By using the q-Cole-Hopf transformation \cite{Nalci et al.}
\begin{equation} u(x,t) = -2 \nu \frac{D_x\phi(x,t)}{\phi(x,t)} ,\label{C H}\end{equation}
where $\phi(x,t)$ is a solution of the q-heat equation (\ref{qheatstandardt}), we find then that
 $u(x,t)$ satisfies the q-Burgers' type evolution equation with cubic nonlinearity
$$\begin{array}{l}\frac{\partial}{\partial t}u(x,t)- \nu D^2_xu(x,t)=\frac{1}{2} \left[(1-M^x_q) u(x,t)D_xu(x,t)\right]-\\ \\\frac{1}{2}\left[D_x\left(u(qx,t)u(x,t)\right)\right]+
\frac{1}{4 \nu} \left[u(q^2x,t)-u(x,qt)\right]u(qx,t)u(x,t), \label{qburger}\end{array}$$
where $M_q^x$ is the delation operator $M_q^x f(x)=f(q x).$
When $q \rightarrow 1$ this equation reduces to the standars Burgers' Equation
\begin{equation}u_t+u u_x= \nu u_{xx}\end{equation}

\subsection{I.V.P. for q-Burgers' Type Equation }
Substituting the operator solution (\ref{IVP}) to (\ref{C H}) we find operator solution for the q-Burgers type equation in the form
\begin{equation} u(x,t) = -2 \nu \frac{e^{\nu t D^2_x } D_x f(x)}{e^{\nu t D^2_x }f(x)} .\label{operator Burgers}\end{equation}This solution
corresponds to the initial function
\begin{equation} u(x,0) = -2 \nu \frac{ D_x f(x)}{
f(x)} .\label{initial Burgers}\end{equation}
Thus, for arbitrary initial value  $u(x,0)=F(x)$  for the q-Burgers equation  we need to solve the initial value problem for the q-heat equation (\ref{qheatstandardt}) with initial
function $f(x)$ satisfying the first order q-difference equation
\begin{equation} (D_x+\frac{1}{2\nu}F(x))f(x)=0.\end{equation}

\section{q-Shock soliton solution}

As a particular solution of the q-heat equation we choose first  \begin{equation}\phi(x,t) = e^{k^2t} e_q\left(kx\right),\end{equation}then we find
solution of the q-Burgers equation as a constant \begin{equation}u(x,t)=-2\nu k. \end{equation}
We notice that for this solution of the q-heat equation, we have an infinite set of zeros, and the space position of  zeros is fixed during time evolution at points $x_n = - q^{n+1}/(q-1)k$, $n = 0,1,...$.

If we choose the linear superposition \begin{equation}\phi(x,t) =  e^{k_1^2t} e_q\left(k_1x\right)+ e^{k_2^2t} e_q\left(k_2x\right),\end{equation}
then we have the q-Shock soliton solution
\begin{equation}u(x,t) = -2\nu \frac{k_1 e^{k_1^2t} e_q\left(k_1x\right)+k_2 e^{k_2^2t} e_q\left(k_2x\right)}{e^{k_1^2t}
e_q\left(k_1x\right)+ e^{k_2^2t} e_q\left(k_2x\right)}.\label{qshock}\end{equation}
This expression is the q-analog of the Burgers shock soliton and for $q\rightarrow 1$ it reduces to the last one.
However, in contrast to the standard Burgers case, due to zeroes of the q-exponential function this expression admits singularities coming from $x$ for some values of parameters $k_1$ and $k_2$.To have regular solution we can follow similar approach as discussed in \cite{Nalci et
al.}, then
 for $k_2 = -k_1$  we have the stationary shock soliton
\begin{equation} u(x,t)=-2\nu k_1 \frac{e_q(k_1 x)-e_q(-k_1 x)}{e_q(k_1 x)+e_q(-k_1 x)}=-2\nu k_1\frac{\sinh_q(k_1 x)}{\cosh_q(k_1 x)} \equiv -2\nu k_1 \tanh_q(k_1 x).\end{equation}
This function has no singularities on the real axis $x$ and everywhere we have regular q-shock soliton as in \cite{Nalci et al.}.However time evolution of shock solitons in \cite{Nalci et al.} produce singularity at finite time. Here we like to find regular in $x$ shock soliton which is regular at any time. \\
 We can choose solution of q-Heat equation (\ref{qheatstandardt}) as
 $$\phi(x,t)=10+e^{k_1^2t} e_q\left(k_1x\right)+ e^{k_2^2t} e_q\left(k_2x\right),$$ then for $k_1 =1$ and $k_2 =-1$  we get the q-Shock soliton
 $$u(x,t)=-2 \nu \frac{e_q(x)-e_q(-x)}{10 e^{-t}+e_q(x)+e_q(-x)}.$$
 This solution describes evolution of shock soliton, so that at $t\rightarrow -\infty$, $u(x,t)\rightarrow 0$, and for   $t\rightarrow \infty$, $u(x,t)\rightarrow -2 \nu \tanh_q x$
In Figures 1,2,3 we plot the regular q-shock soliton for $k_1=1$ and $k_2=-1$ at different time $t=-2,0,5$ with base $q=10$.

\begin{figure}[h]
\begin{center}
\epsfig{figure=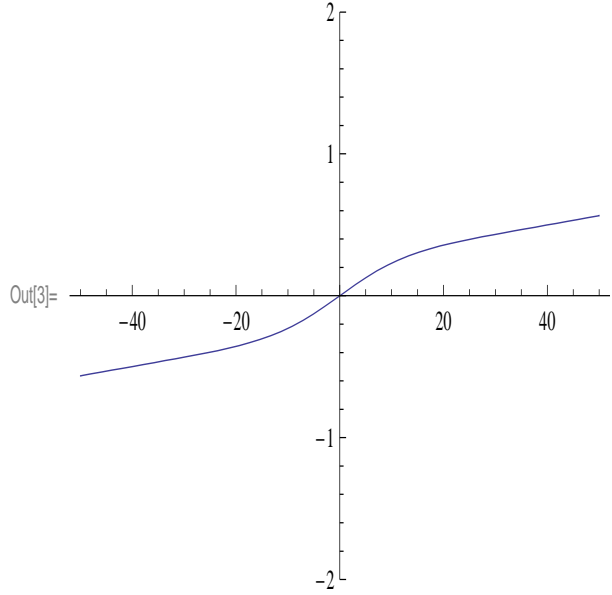,height=8cm,width=8cm}
\end{center}
\caption{q-shock evolution for $\nu=1$, $k_1 = 1$, $k_2 = -1$, $t=-2$ at range (-50, 50)} \label{q-shockevolution1}
\end{figure}

\begin{figure}[h]
\begin{center}
\epsfig{figure=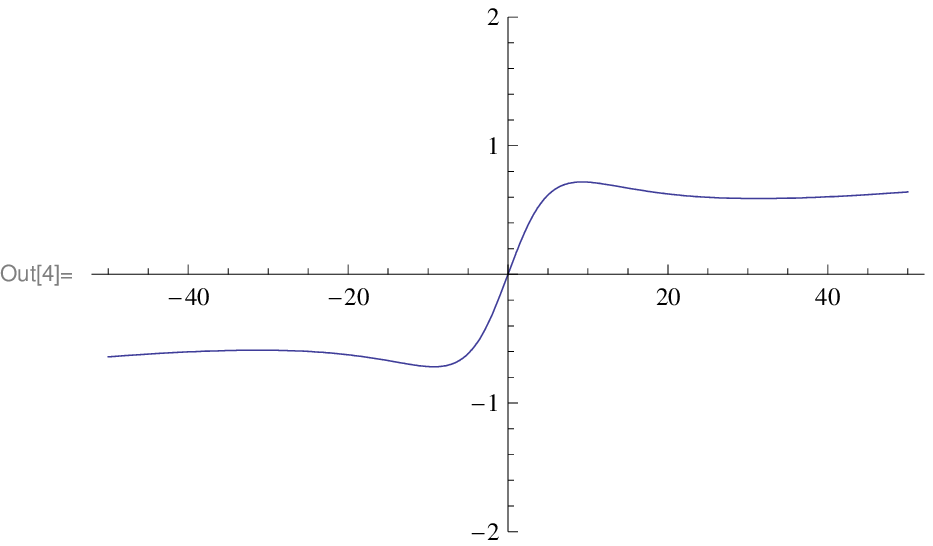,height=8cm,width=8cm}
\end{center}
\caption{q-shock evolution for $\nu=1$, $k_1 = 1$, $k_2 = -1$, $t=0$ at range (-50, 50)} \label{q-shockevolution2}
\end{figure}

\begin{figure}[h]
\begin{center}
\epsfig{figure=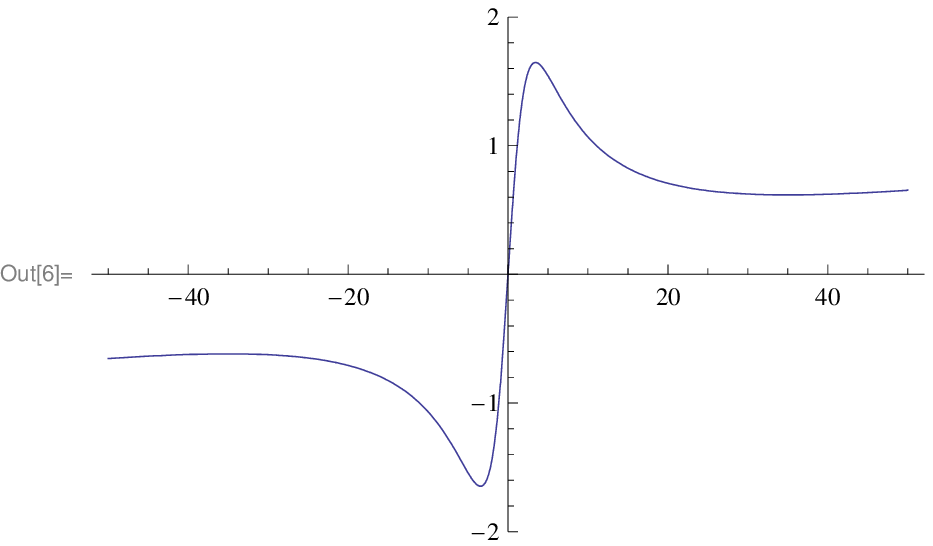,height=8cm,width=8cm}
\end{center}
\caption{q-shock evolution for $\nu=1$, $k_1 = 1$, $k_2 = -1$, $t=5$ at range (-50, 50)} \label{q-shockevolution4}
\end{figure}
 If we plot the regular q-shock soliton evolution for $k_1=1$ and $k_2=-1$ at different ranges of $x$ and with $q=10$, it is remarkable fact that the structure of our q-shock soliton shows self-similar property in the space coordinate $x$. Indeed at the range of parameter $-50< x< 50$,and $-5000< x< 5000$, structure of shock looks almost the same.

For the set of arbitrary numbers $k_1,...,k_N$
\begin{equation}\phi(x,t)= \sum^{N}_{n=1}e^{k_n^2 t} e_q\left(k_nx\right),\end{equation} we have multi-shock
solution in the form
\begin{equation}u(x,t) = -2 \nu\frac{\sum^{N}_{n=1}k_n e^{k_n^2t} e_q\left(k_nx\right)}{\sum^{N}_{n=1}e^{k_n^2t}
e_q\left(k_nx\right)}.\end{equation}

In general this solution admits several singularities.
To have a regular multi-shock solution we can consider the even number of terms $N = 2k$ with opposite wave numbers.
When  $N = 4$ and $k_1 = 1$, $k_2 = -1$,$k_3 = 2$,$k_4 = -2$
we have q-multi-shock soliton solution,

\begin{eqnarray} u(x,t)=-2\nu\frac{\sinh_q(x)+ 2e^{3 t}\sinh_q(2x)}{\cosh_q(x)+ e^{3 t} \cosh_q(2 x)}.\end{eqnarray}

\begin{figure}[h]
\begin{center}
\epsfig{figure=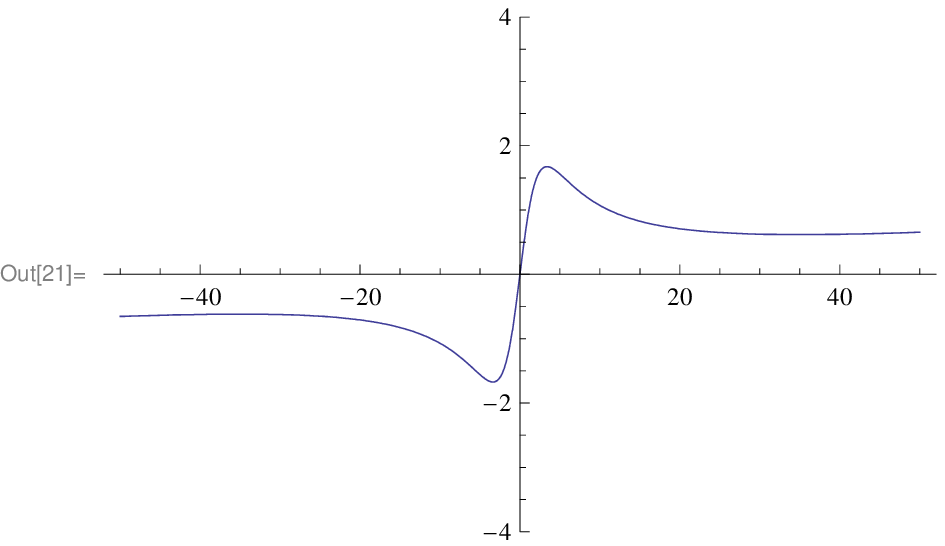,height=8cm,width=8cm}
\end{center}
\caption{q-multi shock evolution for $k_1 = 1$, $k_2 = -1$,$k_3 = 2$, $k_4 = -2$, $t=-10$ and at range (-50, 50)} \label{q-multishockevolution1}
\end{figure}

\begin{figure}[h]
\begin{center}
\epsfig{figure=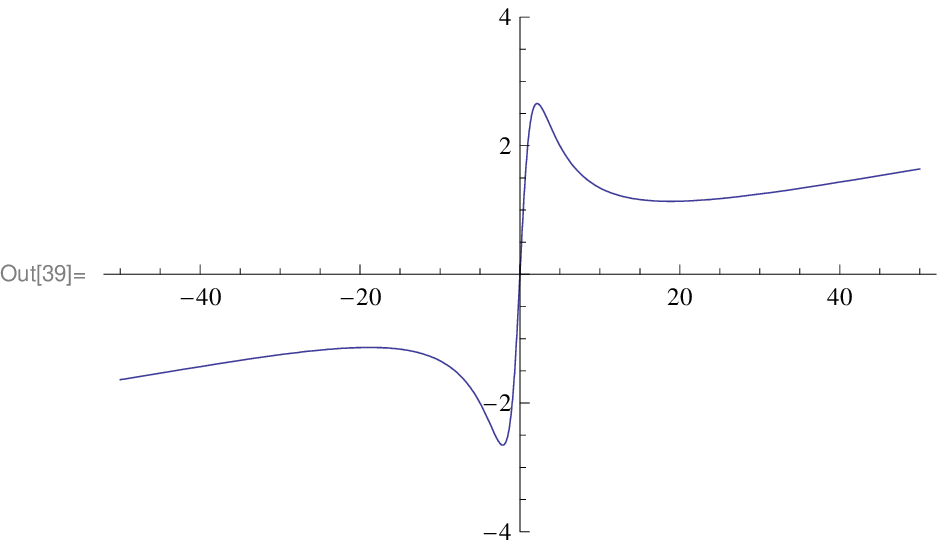,height=8cm,width=8cm}
\end{center}
\caption{q-multi shock evolution for $k_1 = 1$, $k_2 = -1$,$k_3 = 2$, $k_4 = -2$, $t=0$ and at range (-50, 50)} \label{q-multishockevolution2}
\end{figure}

\begin{figure}[h]
\begin{center}
\epsfig{figure=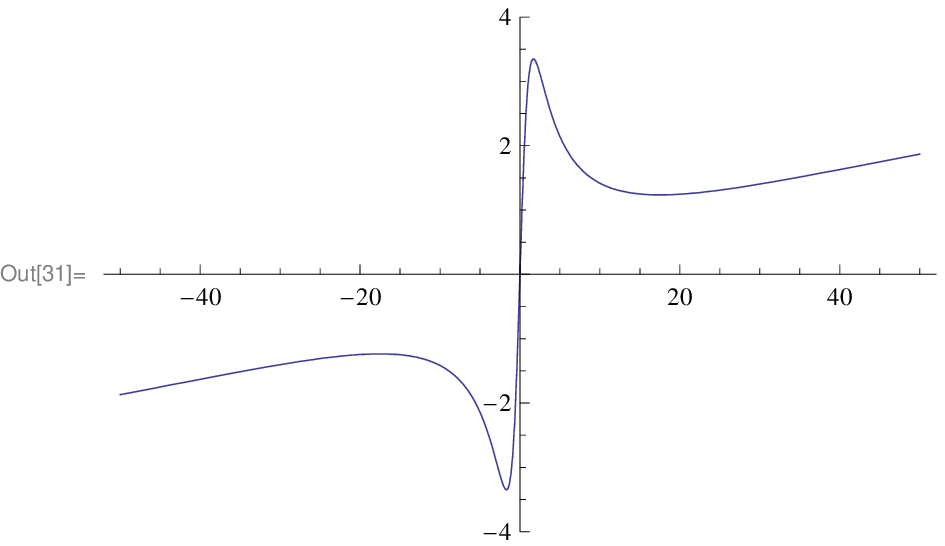,height=8cm,width=8cm}
\end{center}
\caption{q-multi shock evolution for $k_1 = 1$, $k_2 = -1$,$k_3 = 2$, $k_4 = -2$, $t=7$ and at range (-50, 50)} \label{q-multishockevolution3}
\end{figure}

In Figures 7,8,9  we plot $N = 4$ case with values of the wave numbers $k_1 = 1$, $k_2 = -1$, $k_3 = 2$, $k_4 = -2$ at $t=-10,0,7$ and with $q=10$.
This multi-shock soliton is regular everywhere in $x$ for arbitrary time $t$. This result takes place due to absence of zeros for the standard exponential function $e^{k^2 t}$.

If we plot this regular multi q-shock soliton evolution at different ranges of $x$, it is remarkable that the structure of this regular multi q-shock soliton shows also the self-similar property in the space coordinate $x$. Indeed at the range of parameter  $-50< x< 50$, and $-5000< x< 5000$,  structure of multi q-shock looks almost the same.

\section{Standard Time-Dependent q-Schr\"{o}dinger Equation}
The above consideration can be extended to the time dependent Sch\"{o}dinger equation with q-deformed dispersion.
We consider the standard time-dependent q-Sch\"{o}dinger equation
\begin{equation}
\left(\frac{\partial}{\partial t}-\frac{i \hbar}{2 m} D_x^2\right) \psi(x,t)=0 \label{qschrodinger}
\end{equation}
where $\psi(x,t)$ is a complex wave function.\\
One can easily see that $$\psi(x,t)= e^{-\frac{i}{\hbar} \frac{p^2}{2 m } t} e_q\left(\frac{i}{\hbar} p x\right)$$ is the plane wave solution of (\ref{qschrodinger}).
By expanding this in terms of momentum $p$
$$\psi(x,t)= e^{-\frac{i}{\hbar} \frac{p^2}{2 m } t}e_q\left(\frac{i}{\hbar} p x\right)=\sum^\infty_{N=0} (\frac{i}{\hbar})^N \frac{p^N}{[N]_q!} H_N^{(s)}(x,i t;q)$$
we get the set of complex q-Kampe-de Feriet polynomial solutions $$H_N^{(s)}(x,i t;q)= \sum^{[N/2]}_{k=0} \frac{(\frac{i h t}{2 m})^k [N]_q! x^{N-2 k}}{[N-2 k]_q k!}$$ for (\ref{qschrodinger}).

Let us consider the complex version of the q-Cole-Hopf transformation
$$u(x,t)=-\frac{i \hbar}{m}\frac{D_x \psi(x,t)}{\psi(x,t)},$$
then complex velocity function $u(x,t)$ satisfies the complex q-Burgers-Madelung type equation

$$\begin{array}{l}i \hbar \frac{\partial}{\partial t}u(x,t)+\frac{\hbar^2}{2 m} D^2_xu(x,t)=\frac{i \hbar}{2} u(x,t)[1-M_q^x]D_xu(x,t) -\\ \\\frac{i h}{2}\left[D_x\left(u(qx,t)u(x,t)\right)\right]+
\frac{m}{2} \left[u(q^2x,t)-u(x,t)\right]u(qx,t)u(x,t).\end{array}$$
If we write $u=u_1+i u_2$ and separate it into real and imaginary parts , then  we get two fluid model representation where $u_1$ is a Madelung-London-Landau classical velocity, and $u_2$ is the quantum velocity. \\
For the real part we have
\begin{eqnarray}&&-\hbar \frac{\partial}{\partial t} u_2(x,t)+\frac{\hbar^2}{2 m}D_x^2u_1(x,t)=\frac{m}{2}[(u_1(q^2 x,t)-u_1(x,t))(u_1(x,t)u_1(q x,t)- \nonumber \\ \nonumber \\&&u_2(x,t)u_2(q x,t))-(u_2(q^2 x,t)-u_2(x,t))(u_1(x,t)u_2(q x,t)+u_2(x,t)u_1(q x,t))] \nonumber \\ \nonumber \\ &&-\frac{\hbar}{2}[u_1(x,t)[1-M_q^x]D_xu_2(x,t)+u_2(x,t)[1-M_q^x]D_xu_1(x,t)]\nonumber \\ \nonumber \\&&+\frac{\hbar}{2}D_x [u_2(qx,t)u_1(x,t)+u_1(qx,t)u_2(x,t)], \label{realpart}\end{eqnarray}
and for the imaginary part
\begin{eqnarray}&&\hbar \frac{\partial}{\partial t} u_1(x,t)+\frac{\hbar^2}{2 m}D_x^2u_2(x,t)=\frac{m}{2}[(u_1(q^2 x,t)-u_1(x,t))(u_1(x,t)u_2(q x,t)\nonumber \\ \nonumber \\&& +u_2(x,t)u_1(q x,t))+(u_2(q^2 x,t)-u_2(x,t))(u_1(x,t)u_1(q x,t)-u_2(x,t)u_2(q x,t))]+ \nonumber\\ \nonumber \\ && \frac{\hbar}{2}[u_1(x,t)[1-M_q^x]D_xu_1(x,t)-u_2(x,t)[1-M_q^x]D_xu_2(x,t)]-\nonumber \\ \nonumber \\ &&\frac{\hbar}{2}D_x[u_1(q x,t)u_1(x,t)-u_2(q x,t)u_2( x,t)].\label{imaginarypart}\end{eqnarray}

When $q\rightarrow 1$, the real part reduces to the continuity equation $$-(u_2)_t+\frac{\hbar}{2 m}(u_1)_{xx}=(u_1 u_2)_x, $$ and the imaginary part reduces to the Quantum Hamilton-Jacobi equation $$(u_1)_t+\frac{\hbar}{2 m} (u_2)_{xx}=-\frac{1}{2}(u_1^2-u_2^2)_x. $$

For $u_1\equiv v$ and $u_2=-\frac{\hbar}{2 m}(\ln \rho)_x$ where $\rho=|\psi|^2$,\\
the continuity equation is
$$\rho_t + (\rho v)_x=0,$$
and the Euler equation with the quantum potential pressure term is
$$v_t+v v_x=\left( \frac{\hbar^2}{2 m^2}\frac{(\sqrt{\rho})_{xx}}{\sqrt{\rho}}\right)_x.$$

Thus the two fluid system (\ref{realpart}), (\ref{imaginarypart}) is the q-analogue of the coupled q-quantum Hamilton-Jacobi equation and the q-continuity equation.\\
\\
Following similar procedure as in first part of this paper, we can construct particular solutions of our q-Schr\"{o}dinger equation in the form of complex shock solitons. This question is under investigation now.

\end{document}